\documentclass[prb,superscriptaddress,twocolumn,showpacs,preprintnumbers,amsmath,amssymb]{revtex4-1}

\usepackage{graphicx}
\usepackage{epsfig}
\usepackage{dcolumn}
\usepackage{bm}
\usepackage{xspace}
\usepackage[colorlinks=true,citecolor=blue,linkcolor=blue,urlcolor=blue,pdfstartview=FitH]{hyperref}

\def\etbr{$\kappa$-(BEDT\--TTF)$_2$\-Cu\-[N(CN)$_{2}$]Br\xspace}
\def\etcl{$\kappa$-(BEDT\--TTF)$_2$\-Cu\-[N(CN)$_{2}$]Cl\xspace}

\begin{document}


\title{Localized states in the Mott insulator $\kappa$-(BE\-DT\--TTF)$_2$\-Cu\-[N\-(CN)$_{2}$]Cl as probed by photoluminescence}


\author{N. Drichko}
\affiliation{1.\,Physikalisches Institut, Universit\"at Stuttgart, Pfaffenwaldring 57, 70550 Stuttgart, Germany}
\affiliation{Department of Physics and Astronomy, The Johns Hopkins University,  Baltimore, MD 21218, USA}
\author{R. Hackl}
\affiliation{Walther Meissner Institute, Bavarian Academy of Sciences and Humanities, 85748 Garching, Germany}
\author{J. Schlueter}
\affiliation{Materials Science Division, Argonne National Laboratory, Argonne, IL 60439, USA}

\date{\today}

\begin{abstract}
We compare the photoluminescence spectra of the low-temperature Mott insulator \etcl ($T_{\rm MIT}=40$\,K) with spectra of metallic \etbr, which is superconducting below  $T_c=11.8$\,K, in the temperature range between 300 and 20\,K. In the Mott insulating state of \etcl  we observe  a luminescence band at 1.95\,eV due to the recombination of an exciton created by a HOMO-LUMO optical excitation.  This luminescence is quenched both in the high-temperature bad metal state of \etcl and in metallic \etbr.  The observation of the luminescence of an exciton provides evidence for the local character of excitations in the Mott insulating state.
\end{abstract}

\maketitle

\section{Introduction}

The Mott insulator state of solids results from electronic correlations and cannot be derived from band theory in the single-electron approximation \cite{Gebhard}. Mott physics attracts a lot of attention since a Mott insulator state was found in close proximity to superconductivity in, for example,  high-temperature superconductors \cite{Scalapino:1995,Lee:2006,Hanke:2010,Armitage:2010} or organic metals \cite{Toyota07}. Yet, Mott insulators are interesting on their own, since the metal-insulator transition (MIT) occurs without symmetry breaking. The Hubbard model \cite{Hubbard:1963} captures the essential physics and shows directly that the gap opens only due to many-body effects.

It remains  a challenge to find model systems in which the Mott transition can be studied independently of other ordering phenomena such as stripe order in the cuprates, for instance. It is worth mentioning in this regard a recent observation of a Mott insulator to metal transition observed in an  InGaAs/GaAs quantum well, where a control parameter was  the carrier concentration varied via photo-excitation, and no effects of additional ordering are expected \cite{Kappei:2005}. One of the most pure examples of Mott metal-insulator transition in strongly-correlated electron solids are half-filled organic metals, where chemical substitution, pressure or temperature variation are known to induce MITs. The pair \etbr ($\kappa$-ET-Br) and \etcl ($\kappa$-ET-Cl), which belongs to the group of quasi two-dimensional BEDT-TTF-based compounds ( BEDT-TTF stands for bis(ethylenedithio)-tetrathiafulvalene), is a prominent example of the effects of electronic correlations at half-filling\cite{Toyota07,Powell06}. The metallic and insulating properties derive from the conduction band formed by the overlapping highest occupied molecular $\pi$-orbitals (HOMO) of BEDT-TTF. The anion layer is a charge reservoir and, to some extent, determines the geometry of the cation layer. On the average half an electron is removed from the HOMO band of BEDT-TTF, and the intrinsic dimerisation in the $\kappa$-phase layer leads to lattice sites of [(BEDT-TTF)$_2]^{+1}$  and thus a half-filled conduction band.

\begin{figure}
  \centering
    \includegraphics[width=8cm]{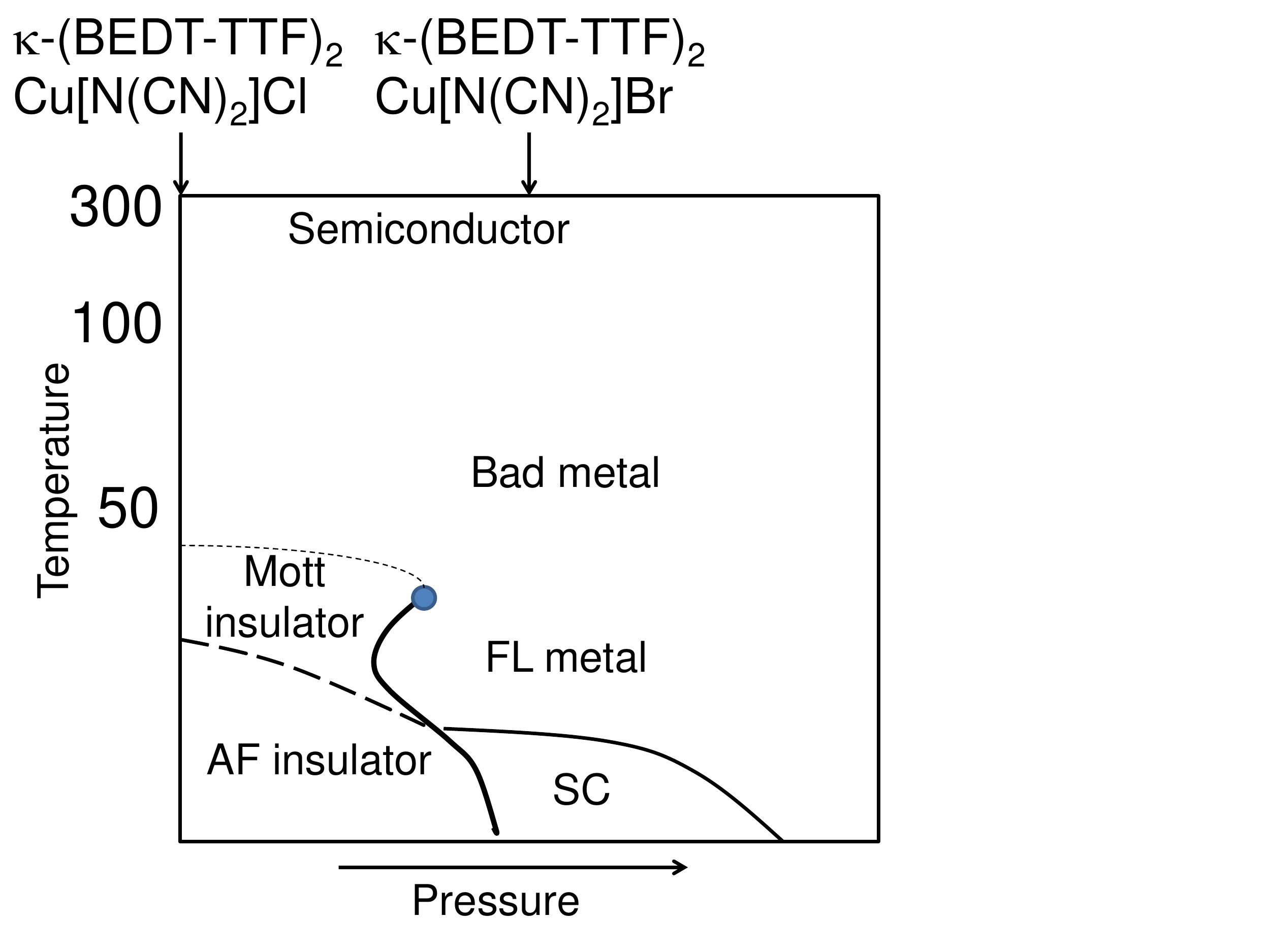}
  \caption{A schematic temperature-pressure phase diagram of the $\kappa$-(BEDT-TTF)$_2$Cu[N(CN)$_2$]$X$ ($X$=Cl, Br) based on the data from Ref.~ \onlinecite{Kagawa09,Yasin2011}. Positions of  of \etbr\ and \etcl\ at ambient pressure are shown by arrows.}
  \label{PD}
\end{figure}

The compounds studied in this work are well characterized by a number of methods, including d.c. resistivity, IR optical properties, and NMR \cite{Kagawa09,Lefebre00,Kagawa05,Yasin2011}. A schematic phase diagram based on the mentioned references is presented in Fig. 1. In $\kappa$-ET-Cl a Mott insulator state manifests itself below the temperature of bad-metal to Mott insulator crossover at about 40\,K, and the compound becomes antiferromagnetic at $T_N=25$\,K.\cite{Kagawa09,Yasin2011} The substitution of Cl by Br in the non-conducting anion layer introduces slight changes in the geometry of the conducting BEDT-TTF layer, which lead to an increase of the transfer integrals between lattice sites and, correspondingly, the bandwidth \cite{Mori99}. The $\kappa$-ET-Br compound shows Fermi-liquid (FL) metallic behavior at low temperatures, and undergoes  a superconducting transition at T$_c$=11.8~K. By comparing $\kappa$-ET-Br and $\kappa$-ET-Cl at various temperatures the bandwidth controlled  MIT can be studied. The electronic structure of these materials is relatively simple, and their physical properties are well described in terms of a  Hubbard model, as a Mott insulator in the case of $\kappa$-ET-Cl, or as a metal close to an MIT in the case of $\kappa$-ET-Br.\cite{Feber}. In agreement with theory, optical infrared investigations show a Mott  gap and an optical transition between the lower and the upper Hubbard band in the Mott insulating state of $\kappa$-ET-Cl. These optical transitions are still observed in the spectra of metallic  $\kappa$-ET-Br, while  a narrow Drude-like carrier response is observed below 100\,K indicating FL behavior.

From an experimental point of view, it is difficult to demonstrate as to whether or not charge is localized, and a material is an insulator at the lowest temperature. For example, the  dc conductivity vanishes only in the zero-temperature limit \cite{Gebhard}. At finite temperatures specific probes of  electronic localization is required. Optical conductivity \cite{Faltermeier07}, as mentioned above, light scattering \cite{Nyhus:1995,Chen:1997,Freericks:2001,Venturini:2002b} or photoluminescence (PL) \cite{Kappei:2005} are possible alternatives. The  inelastic light scattering and  optical conductivity  studies probe the density of states in the occupied bands. The response for the half-filled conductance band is found  below approximately 1 eV for BEDT-TTF-based materials.  PL along with optical absorption at higher frequencies probe interband transitions.  For studies of a metal-insulator transition, PL can be used as a reliable indicator of localization, since in metals PL is weak or absent due to the possibilities of non-radiative relaxation through the metallic bands and screening by free charge carriers. An example of the successful application of this method is a direct observation of a Mott transition in a quantum well \cite{Kappei:2005}. Through PL  an exciton band is observed in the Mott insulator state at low charge carriers concentration, the band disappears at high carrier concentrations in the metallic state.

In this paper, we study PL spectra to investigate the  Mott metal-insulator transition in $\kappa$-ET-Cl achieved by lowering the temperature below 40~K. We observe an exciton luminescence band, which disappears in the high-temperature bad metal state and is absent in the metallic state of  $\kappa$-ET-Br. PL is a local process. Accordingly,  the appearance of a band  in the PL spectra of $\kappa$-ET-Cl below the MIT  demonstrates the local character of the excitations. Further insight can be obtained on the basis of a detailed scheme of electronic levels of $\kappa$-ET-Cl and $\kappa$-ET-Br. The scheme we show compiles data that were previously only found scattered in the literature on the optical transitions in BEDT-TTF-based materials and will be of  use for further electronic Raman scattering studies suggested for this class of compounds \cite{Merino01}.

\section{Experimental}
\label{sec:exp}
High quality single crystals of the $\kappa$-ET-Cl and $\kappa$-ET-Br were grown by standard
electrochemical methods\cite{Williams1990,Kini1990}.  The crystallography and the band structure of these compounds are reviewed in Ref.~\onlinecite{Mori99}. The orthorhombic $a$ and $c$ axes lie in the conducting plane that forms the largest surface of the crystal. The crystals were oriented at room temperature by measuring their infrared spectra and comparing the with the results in Ref.~\onlinecite{Faltermeier07}. The size of the samples used for the studies here was approximately $0.8\times 0.8 \times0.5$\,mm.

The experiments were performed with a calibrated Raman setup equipped with a Jarrell-Ash 25-100 double monochromator having a spectral range from approximately 400 to 723\,nm. The light was detected with a back-illuminated charge coupled device (CCD) cooled to 120\,K. The spectral resolution depends on the wavelength for the fixed slit width used here and improves from 12 to 5.2\,cm$^{-1}$ per mm slit width between 500 and 700\,nm. This change is taken care of in the calibration. In other words, the cross sections shown below correspond to a constant resolution. Minor changes of the line shape may occur in the case of narrow spectral lines but they are irrelevant here.

For excitation we used the Ar$^+$ laser line at 514.5\,nm (2.412\,eV) for both compounds. In order to safely distinguish between inelastic scattering (Raman effect) and PL $\kappa$-ET-Cl was additionally studied with excitation at 476.5\,nm (2.604\,eV). The power impinging on the sample was kept below 1 mW to avoid damage of the samples and to keep the heating in the illuminated spot below 5\,K. All spectra were measured with the scattered photons perpendicular to the incident ones. In most of the cases the incident light was polarized along the $a$ axis. In some cases both polarization were at an angle of $45^{\circ}$ with respect to the $a$ axis. The luminescence turned out to be polarization independent while there are the expected changes on the vibrational Raman-active lines. For the study of the temperature dependence the samples were mounted on the cold finger of a He-flow cryostat having a vacuum of better than $10^{-6}$\,mbar. For attaching the samples on the cold finger we used carbon paint, which does not chemically react with the crystal and  stays sufficiently elastic at low temperature to minimize the stress induced by the differential thermal expansion of the copper holder and the samples.

\section{Results}

\begin{figure}[hbtp]
  \centering
    \includegraphics[width=9cm]{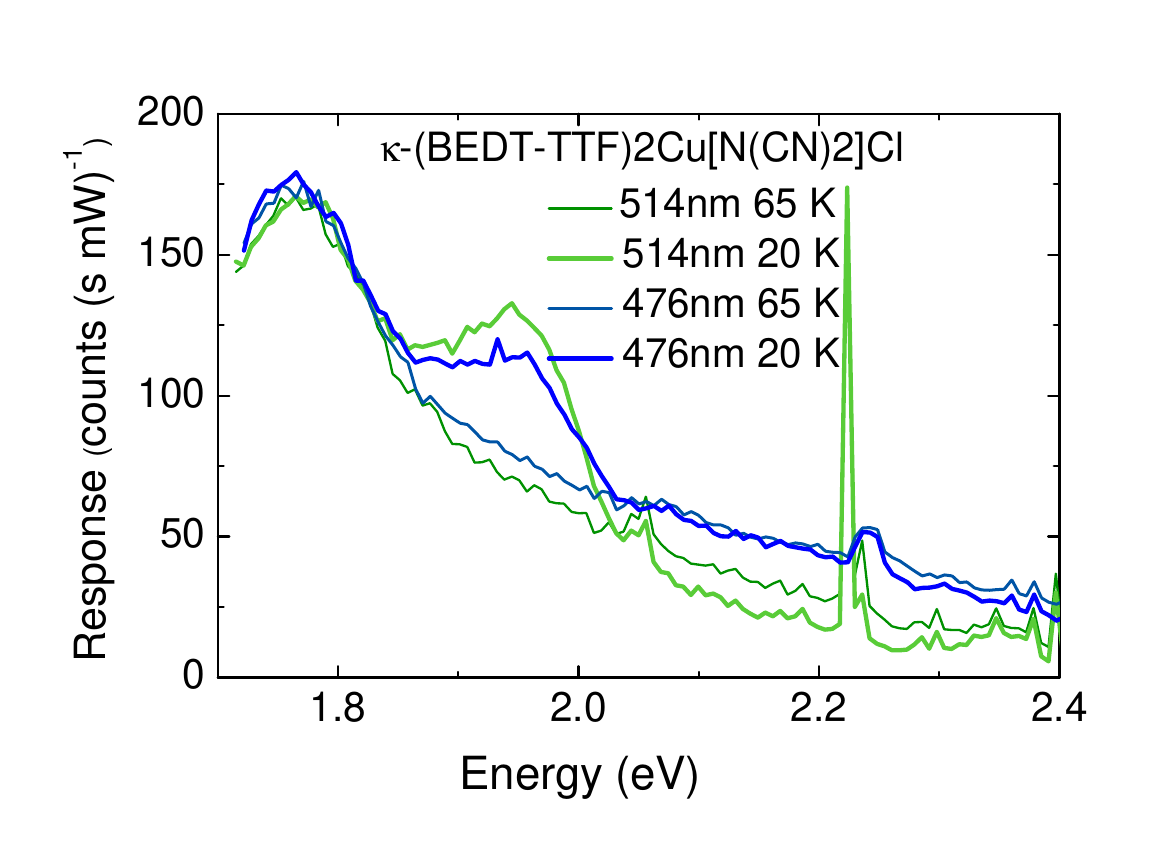}
  \caption{(color online) Photoluminescence spectra of \etcl above and below the MIT (T$_{MIT}$=40~K). The spectra are measured with excitation energies $\hbar\omega_I$ of 2.412 and 2.604\,eV as indicated and plotted as a function of the absolute energy of the detected photons $\hbar\omega_S$. The maxima at 1.76 and 1.95\,eV are independent of the excitation. The narrow lines at 2.22 and 2.43\,eV are Raman lines originating in the C=C stretch
vibration of the two inner rings pairs of the BEDT-TTF molecule at approximately 1502\,cm$^{-1}$ (186\,meV).}
  \label{fig:etcl-E}
\end{figure}

Fig.~\ref{fig:etcl-E} shows the PL spectra of $\kappa$-ET-Cl for the two excitation energies $\hbar\omega_I$ of 2.412 and 2.604\,eV as a function of the scattered photons' energies $\hbar\omega_S$. Two types of excitations are observed: (i) The narrow lines at 2.22 and 2.43\,eV are Raman lines originating in the C=C stretch
vibration of the two inner rings pairs of the BEDT-TTF molecule at $\hbar\Omega = 1502$\,cm$^{-1}$ which would appear at the same shift $\hbar\omega_I - \hbar\omega_S$.\cite{Eldridge96} The difference in the line shape is due to the changes in the resolution and will not be subject of the present paper. (ii) The broad maxima appearing at low temperature at 1.76 and 1.95\,eV  do not vary substantially upon the change of the excitation frequency. Therefore they can be identified as PL. The maximum at 1.95\,eV appears only at low temperature. This data are reproducible
over many cooling cycles and for a number of crystals.

\begin{figure}[hbtp]
  \centering
    \includegraphics[width=8cm]{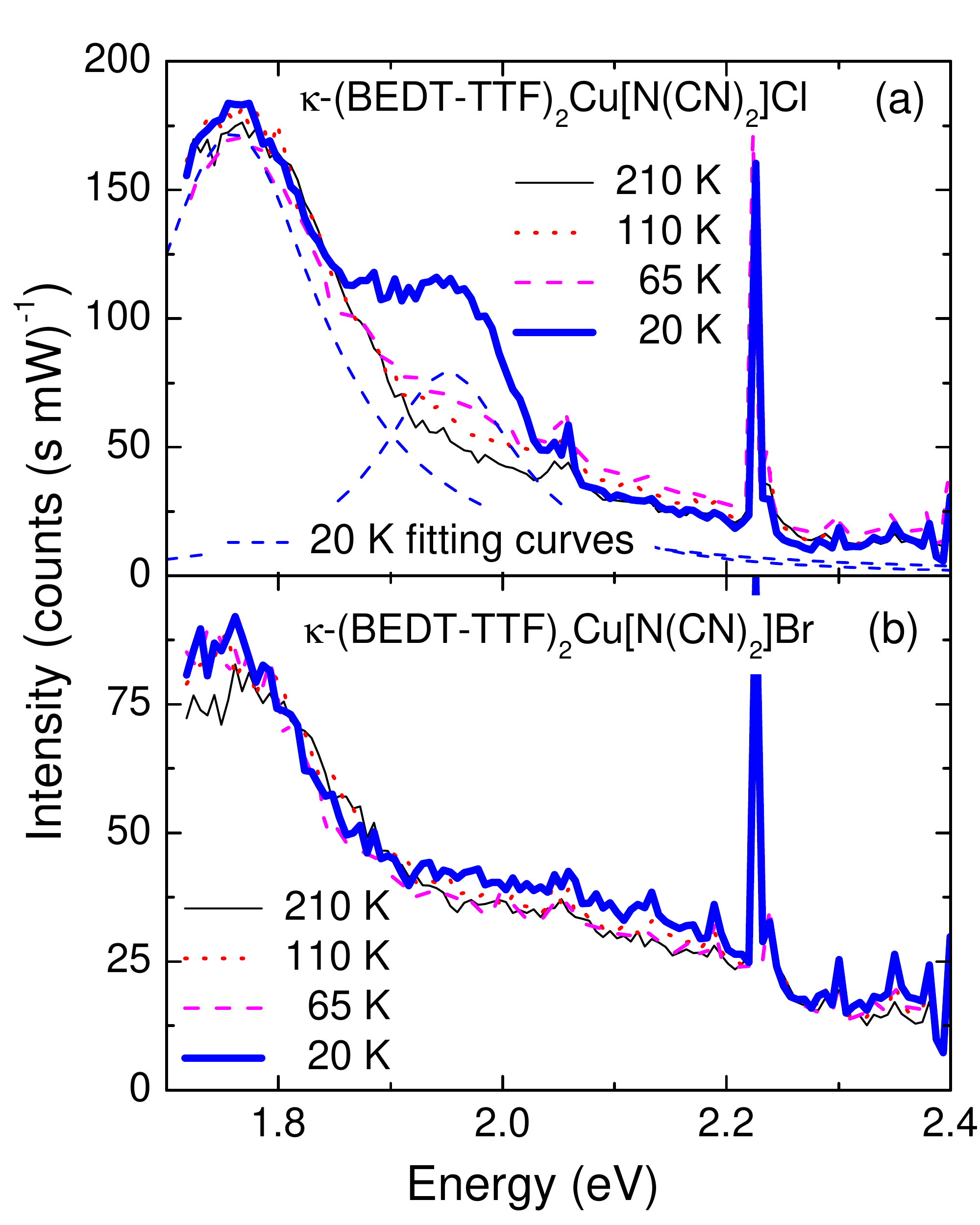}\\
  \caption{(color online) Temperature dependence of photoluminescence spectra of \etcl\ (upper panel) and \etbr (lower panel) for excitation energies 2.41 eV  at 200, 100, 65 and 20 K. The \etcl spectra show an appearance of a new band at 1.95 eV  below T$_{MIT}$.  The upper panel shows a  fit of the spectra at 20 K (dashed lines) with two Lorentz shapes with maxima at 1.76 and 1.95 eV.}\label{fig:T}
\end{figure}

The detailed temperature dependence of the PL spectra of $\kappa$-ET-Cl is plotted in Fig.~\ref{fig:T} (a). The peak at 1.95\,eV is observed in the spectra taken at 20\,K well below the MIT at 40\,K. A minor enhancement of the intensity with respect to that at 200\,K appears already at 100 and 50\,K. The result is reproducible for several cooling cycles and crystals. In contrast, $\kappa$-ET-Br does not show a peak at 1.95\,eV at any temperature (see  Fig.~\ref{fig:T} (b)). The absence of this excitation can now be clearly related to the metallicity of the materials and will be discussed below.

In addition to the maximum at 1.95\,eV,  the spectra of both $\kappa$-ET-Br and $\kappa$-ET-Cl show identical maxima at 1.76\,eV. As demonstrated in Fig.~\ref{fig:T}, these peaks possess the same intensity and position at all the studied temperatures. Obviously, there is no influence of the MIT on this excitation.

\section{Discussion}

The main experimental result of this work is the appearance of an additional luminescence band at 1.95\,eV in the Mott insulator state of $\kappa$-ET-Cl. From a theory point of view one would expect the observation of a low energy gap in the inelastic (Raman) spectra corresponding approximately to the gap between the lower and the upper Hubbard band.\cite{Freericks:2001} We did not find significant indications of this gap in the low-energy range but rather the appearance of the PL transition at 1.95\,eV which will be in the main focus of the following discussion. To this end we look first at the possible transitions in the relevant energy range.

\subsection{Optical transitions in BEDT-TTF-based materials}

The luminescence spectra of other BEDT-TTF-based materials show an intense emission maximum at 1.95\,eV and a weaker one at 1.79\,eV which is interpreted in terms of a vibronic side band \cite{Campos94,Kozlov95}. In this work, the band at 1.76\,eV dominates the spectra and does not depend on temperature. We argue that the band is not  related to the BEDT-TTF layer alone but, rather, has major contributions from the anion layer [Cu(N(CN)$_2$)$X$]$^{-}$ ($X=$\,Cl, Br). In fact, various complexes of copper(I) with nitrogen-based radicals and halogens are known to show a strong luminescence in the visible range \cite{Tronic}. The origin of the transition may depend on the system, and the luminescence can either originate from an exited state of  Cu\,$3d^94s^1$ (Ref.~\onlinecite{Vogler}) or from a metal-ligand transition \cite{Horvath}.  Thus for an  in-depth analysis of the frequency of the luminescence at 1.76\,eV details about the copper coordination in the complex is needed which, however, is beyond of the scope of the present study.

\begin{figure}
\begin{center}
  \includegraphics[width=8cm]{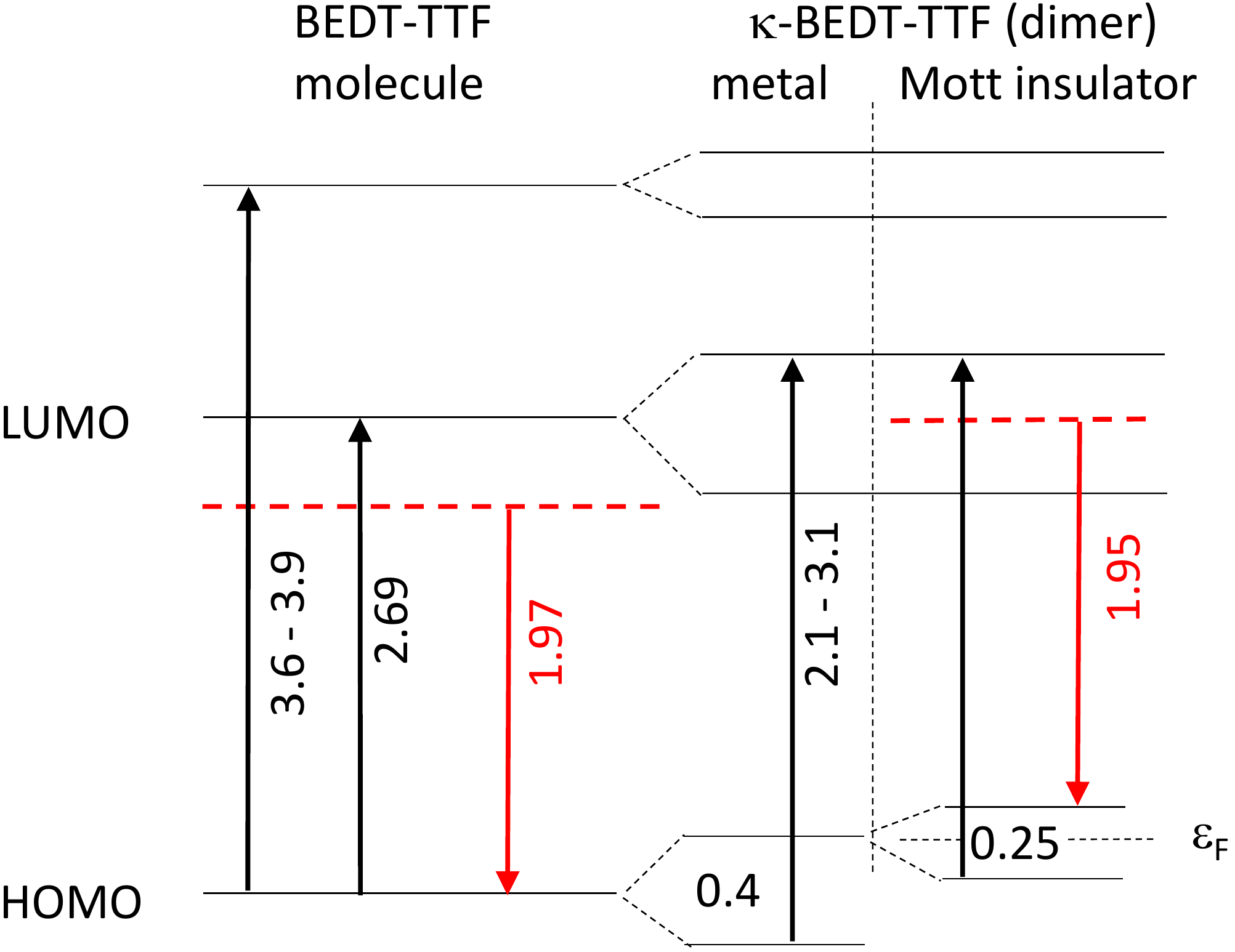}\\
  \end{center}
  \caption{Scheme of the molecular levels of the BEDT-TTF molecule and $\kappa$-phase crystals. The energies are derived from the experimental results of Refs.~\onlinecite{Kozlov95,Truong99,Faltermeier07,Dumm09} and the information about the band structure of the $\kappa$-phase BEDT-TTF-based crystals \cite{Kandpal09}. The left part shows the most prominent levels for the BEDT-TTF molecule. The right part of the scheme presents the energy levels in the $\kappa$-phase crystals. The splitting of the energy levels is associated with the dimerization in the $\kappa$-phase structure. The conduction band formed by HOMO splits into a fully occupied band  and a half-filled one. The size of the splitting  is derived from infrared experiments \cite{Faltermeier07,Dumm09}. A further splitting of the upper half-filled conduction band into a lower and an upper Hubbard band occurs in \etcl\ in the Mott insulating state.}
  \label{fig:AllLevels}
\end{figure}

To interpret our results concerning the temperature behavior of  the band at 1.95\,eV we review the published data on the visible spectra of BEDT-TTF and some BEDT-TTF-based crystals. {\it Ab initio} calculations of the band structure of $\kappa$-ET-Br \cite{Ching97}  and $\kappa$-ET-Cl \cite{Kandpal09} yield information only about the excitations close to the Fermi energy and do not describe optical transitions at and above 2\,eV. Since, to the best of our knowledge, no calculation of molecular levels of BEDT-TTF is published, we need experimental information on  optical transitions of the BEDT-TTF molecule and its salts in a wider energy range.  We  use the absorption, excitation, and luminescence spectra of the BEDT-TTF molecule and its monovalent salt \cite{Kozlov95,Zimmer99,Campos94}, as well as the absorption spectrum of the studied compound\cite{Truong99} for the scheme displayed in Fig.~\ref{fig:AllLevels}.

The left part of the scheme shows results derived from optical spectra of BEDT-TTF molecules. Strong absorption bands are found at 3.58 and 3.86\,eV \cite{Kozlov95} independent of whether the BEDT-TTF molecules or compounds are in solution or crystallized. The changes below 3\,eV are stronger and are related to the formation of BEDT-TTF dimers and/or of electronic bands with considerable dispersion \cite{Kozlov95}. The weak absorption band appearing at 2.6\,eV is usually associated with the HOMO $\rightarrow$ LUMO transition in the monomer. It is very weak if present in the spectra of the isolated molecule and gains intensity in  crystals \cite{Kozlov95,Zimmer99,Sugano87}. For the monovalent salt  of BEDT-TTF studied in Ref.\onlinecite{Kozlov95} or for the compounds studied here, one electron or of half an electron per molecule is removed from the band formed by HOMO orbitals of BEDT-TTF and transferred to the anion. As a result, an additional absorption band is observed at about 2\,eV. This transition is assigned to a  HOMO-1 $\rightarrow$ HOMO interband transition \cite{Kozlov95,Truong99}.

At liquid He temperatures upon excitation with light energies between 2.4 and 4\,eV a luminescence is found at  1.97\,eV  for both neutral BEDT-TTF molecules in solution\cite{Kozlov95} and BEDT-TTF based 3/4-filled crystalline salts. This luminescence is either a LUMO $\rightarrow$ HOMO transition \cite{Campos94} or a bound electron-hole pair \cite{Hummer2004}.

In the crystals of neutral BEDT-TTF \cite{Kozlov95} as well as in $\kappa$-ET-Br\ and $\kappa$-ET-Cl\, the BEDT-TTF molecules are organized in dimers. Here we define dimers as  pairs of face-to-face oriented molecules having a larger overlap of the $\pi$-orbitals. This dimerization of the BEDT-TTF layer leads to the splitting of the bands. For the compounds studied here, the dimerization-induced splitting can be estimated from both the optical conductivity spectra \cite{Faltermeier07} and band structure calculations\cite{Kandpal09} to be 0.4\,eV. In this case,  a formerly 3/4-filled  conductance band (+0.5 per molecule) formed by HOMO orbitals is split  into a completely filled lower band and an upper half-filled conduction band. In $\kappa$-ET-Cl below the MIT at 40\,K, the conduction band further splits into a lower fully occupied and upper unoccupied Hubbard band (see r.h.s. of Fig.~\ref{fig:AllLevels}). In Ref.~\onlinecite{Kozlov95} the slight difference between the luminescence of the BEDT-TTF molecule and that of the neutral crystal (1.97~eV vs. 1.96~eV) was assigned to the splitting due to the dimerization. However, the estimates done above show, that this difference is much smaller than the splitting.

\subsection{Free and bound electron-hole pairs in the Mott insulator state}

Intrinsic exciton luminescence  is  observed for many molecular crystals formed by aromatic molecules \cite{Hochstrasser62,Hummer2004,bala2006}. Molecular crystals such as  neutral  crystalline BEDT-TTF are band insulators, which explains the presence of  the luminescence in the latter.

In the low-temperature Mott insulating state of $\kappa$-ET-Cl we observe a photoluminescence at 1.95\,eV excited by photons with energies of either 2.412 or 2.604\,eV. This excitation across the HOMO-LUMO gap creates an exciton with relatively high energy and a small radius (Frenkel exciton). An observation of Frenkel excitons in insulating and semiconducting organic aromatic compounds is fairly common with anthracene being a good example \cite{Hummer2004}. Therefore, the explanation of PL spectra in BEDT-TTF-based materials in terms of recombining excitons is probably the most natural approach.

The exciton PL band at 1.95\,eV is absent in the spectra of the the high-temperature bad-metal and the low-temperature metallic state of $\kappa$-ET-Br, as well as for $\kappa$-ET-Cl at temperatures above 40\,K. In the metallic and bad-metal states the created electron-hole pair does not form an exciton due to the high enough screening values\cite{Nakamura09} in a system of quasi-free charge carriers. In the Mott insulating state of $\kappa$-ET-Cl the exciton is observed in a PL process because in this case the screening is essentially reduced.

Controlling the luminescence in solids is an important applied problem \cite{Mutai2005}. Our results show that a transition from a Mott insulating state into a metallic state  is a viable way towards quenching luminescence indicating that electronic correlations can provide functionality in organic conductors.

The Mott transition paves the way towards other instabilities such as phase separation \cite{Emery:1990,Grilli:1991} or anti-ferromagnetism. Thus, there may be a cascade of transitions in $\kappa$-ET-Cl triggered by the MIT, for instance, ferro-electricity as observed recently below approximately 30\,K (Ref. \onlinecite{Lunkenheimer:2012}).

\section{Conclusions}

We studied the photoluminescence spectra of the organic conductors \etbr\ and \etcl\ in the spectral region of 1.7-2.6\,eV at temperatures between 20 and 300\,K using laser radiation at 2.412 and 2.604\,eV.  For \etcl\ below $T=40$\,K we observe the appearance of a band at 1.95\,eV which we assign to the luminescence decay of an exciton right below the HOMO-LUMO gap (dashed horizontal line in Fig.~\ref{fig:AllLevels}). This luminescence is quenched by screening in the metallic state of \etbr\ as well as in the high-temperature bad metal state in both materials. The presence of the exciton luminescence in the Mott insulating state demonstrates the local properties of this ground state and is an example of how luminescence can be controlled by electronic correlations.

\section{Acknowledgements}

We are indebted to N. P. Armitage and M. Dressel for stimulating discussions.
N.D. acknowledges support by the German Research Foundation (DFG) (grant-no. DR228/39), by the Margarete von Wrangell Habilitationstipendium; work at JHU was supported  by the H. Blewett Fellowship of the American Physical Society and by DOE grant for The Institute of Quantum Matter DE-FG02-08ER46544. R.H. acknowledges support by the DFG via the Collaborative Research Center TRR\,80. Work supported by UChicago Argonne, LLC, Operator of Argonne National Laboratory ("Argonne"). Argonne, a U.S. Department of Energy Office of Science laboratory, is operated under Contract No. DE-AC02-06CH11357

\bibliography{./literatureR}

\end{document}